\documentclass[10pt, a4paper, twocolumn, twoside]{IEEEtran}

\usepackage{cite}
\usepackage{enumerate}
\usepackage{graphicx}
\usepackage[cmex10]{amsmath} 
\usepackage{amssymb}
\usepackage{subfig}
\usepackage{xspace}
\usepackage[lined,linesnumbered,ruled,commentsnumbered]{algorithm2e} 
\usepackage{colonequals}
\usepackage[mathscr]{euscript}
\usepackage{amsthm}
\usepackage{mathrsfs}

\usepackage{epsfig}
\usepackage{epstopdf}

\usepackage{tikz}
\usepackage{pgfplots}
\usepackage{pgfplotstable}

\usetikzlibrary{arrows,shapes,backgrounds,plotmarks,positioning,automata,calc,trees}

\pgfplotsset{compat=newest}                         
\pgfplotsset{plot coordinates/math parser=false}
\newlength\figureheight
\newlength\figurewidth

\newtheorem{theorem}{Theorem}[section]
\newtheorem{lemma}[theorem]{Lemma}

\newtheorem{corollary}[theorem]{Corollary}

\newtheorem{example}[theorem]{Example}

\newcommand{\Real}{\mathbb{R}}
\newcommand{\RealP}{\mathbb{R}_{+}}    
\newcommand{\Z}{\mathbb{Z}}            
\newcommand{\ZP}{\mathbb{Z}_{+}}            
\newcommand{\F}{\mathbb{F}}            

\newcommand{\op}{\text}
\newcommand{\RZ}[1]{\mathsf{Z}_{#1}}
\newcommand{\RW}[1]{\mathsf{W}_{#1}}
\newcommand{\Rz}[1]{\mathsf{z}_{#1}}

\newcommand{\RRCO}{\mathscr{R}_{\op{CO}}}
\newcommand{\RRACO}{\mathscr{R}_{\op{ACO}}^*}
\newcommand{\RRNCO}{\mathscr{R}_{\op{NCO}}^*}
\newcommand{\RACO}{R_{\op{ACO}}}
\newcommand{\RNCO}{R_{\op{NCO}}}
\newcommand{\Card}[1]{|#1|}
\newcommand{\Set}[1]{\{#1\}}

\newcommand{\Pat}{\mathcal{P}}

\newcommand{\SFM}{\text{SFM}}

\newcommand{\rv}{\mathbf{r}}         

\newcommand{\FuHash}[1]{f_{#1}^{\#}}
\newcommand{\FuHashHat}[1]{\hat{f}_{#1}^{\#}}

\begin{document}

\title{A Practical Approach for Successive Omniscience}

\author{Ni~Ding, Rodney~A.~Kennedy and Parastoo~Sadeghi
\thanks{\IEEEauthorblockA{The authors are with the Research School of Engineering, College of Engineering and Computer Science, the Australian National University (email: $\{$ni.ding, rodney.kennedy, parastoo.sadeghi$\}$@anu.edu.au)}.}
}


\maketitle
\thispagestyle{empty}
\pagestyle{empty}

\begin{abstract}
The system that we study in this paper contains a set of users that observe a discrete memoryless multiple source and communicate via noise-free channels with the aim of attaining omniscience, the state that all users recover the entire multiple source. We adopt the concept of successive omniscience (SO), i.e., letting the local omniscience in some user subset be attained before the global omniscience in the entire system, and consider the problem of how to efficiently attain omniscience in a successive manner. Based on the existing results on SO, we propose a CompSetSO algorithm for determining a complimentary set, a user subset in which the local omniscience can be attained first without increasing the sum-rate, the total number of communications, for the global omniscience. We also derive a sufficient condition for a user subset to be complimentary so that running the CompSetSO algorithm only requires a lower bound, instead of the exact value, of the minimum sum-rate for attaining global omniscience. The CompSetSO algorithm returns a complimentary user subset in polynomial time. We show by example how to recursively apply the CompSetSO algorithm so that the global omniscience can be attained by multi-stages of SO.
\end{abstract}

\section{introduction}

The problem of communication for omniscience (CO) was originally formulated in \cite{Csiszar2004}. It is assumed that there are a finite number of users in a system. Each of them observes a distinct component of a discrete multiple correlated source in private. The users are allowed to exchange their observations over public authenticated noiseless broadcast channels. The purpose is to attain \textit{omniscience}, the state that each user obtains all the components in the entire multiple source. The CO problem in \cite{Csiszar2004} is based on an \textit{asymptotic model} where the length of observation sequence is allowed to approach infinity. Whereas the finite linear source model \cite{Chan2011ITW} and packet model in the coded cooperative data exchange (CCDE) \cite{Roua2010,SprintRand2010,CourtIT2014} can be considered as the \textit{non-asymptotic model} where the number of observations is finite and the communication rates are restricted to be integral.

While, in the majority of the studies, e.g. \cite{Csiszar2004,CourtIT2014,MiloIT2015}, the omniscience is attained in a one-off manner, the concept of successive omniscience (SO) is proposed in \cite{ChanSuccessiveIT,Ding2015NetCod}. The idea is to let the omniscience be achieved in a successive manner: attain the local omniscience in a user subset before the global omniscience. It is shown in \cite{ChanSuccessiveIT} that we can attain the local omniscience in a \textit{complimentary} set, a user subset that has a multivariate mutual information (MMI) no less than the MMI in the entire system, while still keep the overall communication rates for CO minimized.
SO is also an attractive idea when we want to design a practical method for CO. Firstly, solving the local omniscience problem (in a user subset) is less complex than the global one. Secondly, in a large scale system where the global omniscience takes a long time, it is practical to let a small group of users attain the local omniscience first so that they can be treated as a super user, which reduces the dimension of the CO problem. Thirdly, when the system parameters vary, e.g., when some users in the original system move out of the communication range,\footnote{This could happen in a large scale CCDE system where the users are mobile clients.} the solution to SO is optimal up-to-the-date.

While the implementation of SO boils down to the problem of how to determine a complimentary user subset, the study in \cite{ChanSuccessiveIT} does not provide a practical solution and the iterative merging algorithm proposed in \cite{Ding2015NetCod} only applies to the CCDE systems. In this paper, we propose an efficient algorithm for searching the complimentary subset for SO in both asymptotic and non-asymptotic models. First, the necessary and sufficient condition for a user subset to be complimentary in \cite{ChanSuccessiveIT} is converted to the one that is conditioned on the value of the Dilworth truncation. Based on this condition, we propose an algorithm for searching the complimentary user subset for SO, which is called the CompSetSO algorithm. While running this algorithm still requires the value of the minimum sum-rate in both asymptotic and non-asymptotic models, we derive a suffient condition for a user subset to be complimentary so that knowing only the lower bound on the minimum sum-rate is sufficient. We show that, based on this lower bound, the CompSetSO algorithm either searches a complimentary subset or returns a global-omniscience-achievable rate vector with the minimum sum-rate in $O(|V| \cdot \SFM(|V|))$ time. Here, $V$ denotes the user set and $\SFM(|V|)$ denotes the complexity of minimizing a submodular set function that is defined on $2^V$. Finally, an example is presented to show how to attain omniscience in multi-stages of SO in a CCDE system by adopting random linear network coding (RLNC) scheme \cite{Ho2006RLNC}.

\section{System Model}
\label{sec:system}

Let $V$ with $\Card{V}>1$ be a finite set that contains the indices of all users in the system. We call $V$ the \textit{ground set}. Let $\RZ{V}=(\RZ{i}:i\in V)$ be a vector of discrete random variables indexed by $V$. For each $i\in V$, user $i$ privately observes an $n$-sequence $\RZ{i}^n$ of the random source $\RZ{i}$ that is i.i.d.\ generated according to the joint distribution $P_{\RZ{V}}$. We allow the users exchange their sources directly so as to let all users in $V$ recover the source sequence $\RZ{V}^n$. The state that each user obtains the total information in the entire multiple source is called \textit{omniscience}, and the process that users communicate with each other to attain omniscience is called \textit{communication for omniscience (CO)} \cite{Csiszar2004}. For $X \subseteq V$, the \textit{(local) omniscience} in $X$ refers to the state that all users $i \in X$ recover the sequence $\RZ{X}^n$. The \textit{global omniscience} in the ground set $V$ is a special case of the local omniscience when $X = V$.

\subsection{Minimum Sum-rate}

Consider the local omniscience problem in $X \subseteq V$. Let $\rv_X=(r_i:i\in X)$ be a rate vector indexed by $X$. We call $\rv_X$ an \textit{achievable rate vector} if the omniscience in $X$ can be attained by letting users communicate with the rates designated by $\rv_X$. Let $r$ be the function associated with $\rv_X$ such that
$$ r(C)=\sum_{i\in C} r_i, \quad \forall C \subseteq X $$
with the convention $r(\emptyset)=0$. We call $r(X)$ the \textit{sum-rate} of $\rv_X$. For $C \subseteq X$, let $H(\RZ{C})$ be the amount of randomness in $\RZ{C}$ measured by Shannon entropy and $H(\RZ{X \setminus C}|\RZ{C})=H(\RZ{X})-H(\RZ{C})$ be the conditional entropy of $\RZ{X \setminus C}$ given $\RZ{C}$. In the rest of this paper, without loss of generality, we simplify the notation $\RZ{C}$ by $C$.

It is shown in \cite{Csiszar2004} that an achievable rate vector $\rv_X$ must satisfy the Slepian-Wolf (SW) constraints: $r(C) \geq H(C|X \setminus C), \forall C \subset X$. The interpretation is: To attain the omniscience in $X$, the total amount of information sent from user set $C$ should be at least equal to what is missing in $X \setminus C$. We have the set of all achievable rate vectors being
$$ \RRCO(X)=\Set{ \rv_X\in\Real^{|X|} \colon r(C) \geq H(C|X\setminus C),\forall C \subset X }.$$

In an asymptotic model, we study the CO problem by considering the asymptotic limits as the \emph{block length} $n$ goes to infinity so that the communication rates could be real or fractional; In a non-asymptotic model, the block length $n$ is restricted to be finite and the communication rates are required to be integral. $\RACO(X)=\min\Set{r(X) \colon \rv_X\in \RRCO(X)}$ and $\RNCO(X)=\min\Set{r(X) \colon \rv_X\in \RRCO(X) \cap \Z^{|X|} }$ are the minimum sum-rates for attaining omniscience in $X$ in the asymptotic and non-asymptotic models, respectively. $\RRACO(X)=\Set{\rv_X\in \RRCO(X) \colon r(X)=\RACO(X)} $ and $\RRNCO(X)=\Set{\rv_X \in \RRCO(X) \cap \Z^{|X|} \colon r(X)=\RNCO(X)}$ are the corresponding optimal rate vector sets for the asymptotic and non-asymptotic models, respectively.

For $X \subseteq V$, let $\Pi(X)$ be the set of all partitions of $X$ and $\Pi'(X) = \Pi(X) \setminus \Set{X}$. It is shown in \cite{Csiszar2004,ChanMMI,Ding2016IT} that
    \begin{equation} \label{eq:MinSumRate}
            \RACO(X) = \max_{\Pat \in \Pi'(X)} \sum_{C \in \Pat} \frac{H(X) - H(C)}{|\Pat|-1}
    \end{equation}
and $\RNCO(X) = \lceil \RACO(X) \rceil$.
The maximization problem \eqref{eq:MinSumRate} can be solved and an optimal rate vector in $\RRACO(X)$ or $\RRNCO(X)$ can be determined in $O(|X|^2 \cdot \SFM(|X|))$ time \cite{Ding2016IT}.\footnote{$O(\SFM(|X|))$ is the complexity of solving the problem $\min\Set{f(C) \colon C \subseteq X}$ for submodular set function $f$, which is strongly polynomial \cite{Fujishige2005}.}

\begin{example} \label{ex:aux}
Consider the system where $V=\{1,\dotsc,5\}$ and each user observes respectively
    \begin{equation}
        \begin{aligned}
            \RZ{1} & = (\RW{a},\RW{b},\RW{c},\RW{d},\RW{f},\RW{g},\RW{i},\RW{j}),  \\
            \RZ{2} & = (\RW{a},\RW{b},\RW{c},\RW{f},\RW{i},\RW{j}),\\
            \RZ{3} & = (\RW{e},\RW{f},\RW{h},\RW{i}),   \\
            \RZ{4} & = (\RW{b},\RW{c},\RW{e},\RW{j}), \\
            \RZ{5} & = (\RW{b},\RW{c},\RW{d},\RW{h},\RW{i}),
        \end{aligned}  \nonumber
    \end{equation}
where $\RW{j}$ is an independent uniformly distributed random bit. In the corresponding CCDE system, each $\RW{j}$ represents a packet and $\RZ{i}$ denotes the packets received by mobile client $i$ from a base station. The mobile clients in $V$ transmit linear combinations of $\RZ{i}$s by some network coding scheme, e.g., \cite{Roua2010}, over noiseless peer-to-peer channels in order to recover all packets in $\RZ{V}$. In this system, $\RACO(V)= \frac{13}{2}$ and $\RNCO(V) = 7$. We have $(\frac{9}{2},0,\frac{1}{2},\frac{1}{2},1) \in \RRACO(V)$ and $(5,0,1,1,0) \in \RRNCO(V)$ being one of the optimal rate vectors in the asymptotic and non-asymptotic models, respectively.\footnote{The optimal rate vector is not unique, i.e., $\RRACO(V)$ and $\RRNCO(V)$ are not singleton, in general.}
\end{example}


\section{Successive Omniscience}

The idea of successive omniscience (SO) is proposed in \cite{ChanSuccessiveIT,Ding2015NetCod}, which allows the omniscience to be achieved in a successive manner: first attain the local omniscience in a user subset $X$; then solve the global omniscience problem in $V$ by assuming that all the users $i \in V \setminus X$ have obtained the information in the communications for achieving the local omniscience in $X$.

\subsection{Complimentary User Subset}

Let $X \subseteq V$ such that $|X| > 1$ be a non-singleton user subset. $X$ is called a \textit{complimentary} subset (for SO) if the local omniscience in $X$ can be achieved first without increasing the minimum sum-rate, $\RACO(V)$ and $\RNCO(V)$ in the asymptotic and non-asymptotic models, respectively, for the global omniscience in $V$\cite{ChanSuccessiveIT}.
To be more specific, take the asymptotic model for example. For a non-singleton subset $X \subseteq V$, let $\rv_X \in \RRACO(X)$ be an optimal rate vector for attaining the local omniscience in $X$. Let $\rv'_V = (r'_i \colon i \in V )$ where $r'_i = r_i$ if $i \in X$ and $r'_i=0$ otherwise. If there exists a rate vector $\rv''_V \in \RealP^{|V|}$ such that $\rv'_V + \rv''_V \in \RRACO(V)$, then local omniscience in $X$ is complimentary. Similarly, in the non-asymptotic model, let $\rv'_V $ be constructed in the same way by $\rv_X \in \RRNCO(X)$ for $X \subseteq V$. $X$ is complimentary if there exists a rate vector $\rv''_V \in \ZP^{|V|}$ such that $\rv'_V + \rv''_V \in \RRNCO(V)$.

\begin{example}\label{ex:auxSO}
    Consider the asymptotic model for the system in Example~\ref{ex:aux}. The minimum sum-rate for achieving the local omniscience in $\Set{1,2}$ is $\RACO(\Set{1,2}) = 2$ and $\rv_{\Set{1,2}} = (r_1,r_2) = (2, 0) \in \RRACO(\Set{1,2})$ is an optimal rate vector. In this case, we have $\rv'_V = (2,0,0,0,0)$. After the users transmit $\rv'_V$ for achieving the local omniscience in $\Set{1,2}$, we can let them transmit $\rv''_V = (\frac{5}{2},0,\frac{1}{2},\frac{1}{2},1)$, which attains the global omniscience in $V = \Set{1,\dotsc,5}$. Therefore, $\Set{1,2}$ is a complimentary user subset. Here, $\rv'_V + \rv''_V = (\frac{9}{2},0,\frac{1}{2},\frac{1}{2},1) \in \RRACO(V)$ is an optimal rate vector as shown in Example~\ref{ex:aux}. In other words, the optimal rate vector $(\frac{9}{2},0,\frac{1}{2},\frac{1}{2},1)$ for the omniscience in $V = \Set{1,\dotsc,5}$ can be implemented in an SO manner so that the local omniscience in $\Set{1,2}$ can be achieved before the global omniscience.

    For the non-asymptotic model, $\Set{1,2}$ is also a complimentary user subset since SO can be done by rate vectors $\rv'_V = (2,0,0,0,0)$ and $\rv''_V = (3,0,1,1,0)$, which first achieve local omniscience in $\Set{1,2}$ and then global omniscience in $V = \Set{1,\dotsc,5}$. Here, $\rv'_V + \rv''_V = (5,0,1,1,0) \in \RRNCO(V)$ is an optimal rate vector as shown in Example~\ref{ex:aux}.
\end{example}

\subsubsection{necessary and sufficient condition}

The necessary and sufficient condition for a user subset $X$ to be complimentary is derived in \cite{ChanSuccessiveIT} for both asymptotic and non-asymptotic models, which is stated as follows.

\begin{theorem}[necessary and sufficient condition {\cite[Theorems 4.2 and 5.2]{ChanSuccessiveIT}}]\label{theo:SO}
    In an asymptotic model, a non-singleton user subset $X \subset V$ is complimentary if and only if $H(V) - H(X) + \RACO(X) \leq \RACO(V)$; In a non-asymptotic model, a non-singleton user subset $X \subset V$ is complimentary for SO if and only if $H(V) - H(X) + \RNCO(X) \leq \RNCO(V)$.\footnote{Let $I(X)$ denote the multivariate mutual information (MMI) in $X$. In {\cite[Theorems 4.2 and 5.2]{ChanSuccessiveIT}}, the necessary and sufficient condition for $X$ to be complimentary is $I(X) \geq I(V)$ and $\lfloor I(X) \rfloor \geq \lfloor I(V) \rfloor$ for the asymptotic and non-asymptotic models. They can be converted to the conditions in Theorem~\ref{theo:SO} via the dual relationships: $\RACO(V) = H(V)- I(V) $ and $\RNCO(V) = H(V)- \lfloor I(V) \rfloor$ \cite{ChanMMI,Court2011}.
    Also, since the ground set $V$ is always a complimentary subset, we restrict our attention to the non-singleton proper subsets $X$ of $V$ that are complimentary.} \hfill\IEEEQED
\end{theorem}

Here, $H(V) - H(X)$ is the amount of information that is missing in user subset $X$, the omniscience of which only relies on the transmissions from the users in $V \setminus X$. If we let the users in $X$ attain local omniscience with the minimum sum-rate $\RACO(X)$, the users in $V \setminus X$ are required to transmit at least $H(V) - H(X)$ for attaining the global omniscience. Then, the total number of transmissions is no less than $H(V) - H(X) + \RACO(X)$. If $H(V) - H(X) + \RACO(X)>\RACO(V)$, the global omniscience is not achievable by the minimum sum-rate $\RACO(V)$ if we allow the users in $X$ to attain the local omniscience first. The condition $H(V) - H(X) + \RNCO(X) \leq \RNCO(V)$ for the non-asymptotic model in Theorem~\ref{theo:SO} can be interpreted in the same way.

However, Theorem~\ref{theo:SO} cannot be directly applied for determining a complimentary subset since the power set $2^V$ is exponentially large in $|V|$. In the following context, we convert Theorem~\ref{theo:SO} to the conditions on the Dilworth truncation and propose a polynomial time algorithm for searching a complimentary user subset for SO.

For $0 \leq \alpha \leq H(V)$, let
$$ \FuHash{\alpha}(X) = \begin{cases}
                       0, & \mbox{if $X = \emptyset$} \\
                       \alpha - H(V) + H(X), & \mbox{otherwise}.
                     \end{cases}$$
$\FuHashHat{\alpha}(X) = \min_{\Pat \in \Pi(X)} \sum_{C \in \Pat} \FuHash{\alpha}(C), \forall X \subseteq V$ is the \textit{Dilworth truncation} of $\FuHash{\alpha}$ \cite{Dilworth1944}.

\begin{corollary} \label{coro:SOEQV1}
    In an asymptotic model, a non-singleton user subset $X \subset V$ is complimentary for SO if and only if $\FuHash{\RACO(V)}(X) = \FuHashHat{\RACO(V)}(X)$; In a non-asymptotic model, a non-singleton user subset $X \subset V$ is complimentary for SO if and only if $\FuHash{\RNCO(V)}(X) = \FuHashHat{\RNCO(V)}(X)$. \hfill \IEEEQED
\end{corollary}

The proof of Corollary~\ref{coro:SOEQV1} is in Appendix~\ref{app:coro:SOEQV1}

\begin{example} \label{ex:auxSOCompSet}
    For the system in Example~\ref{ex:aux}, we have
     \begin{equation}
        \begin{aligned}
            &\Set{X \subset V \colon |X| > 1, H(V) - H(X) + \RACO(X) \leq \RACO(V)} \\
            &\qquad = \Set{X \subset V \colon |X|>1, \FuHash{\RACO(V)}(X) = \FuHashHat{\RACO(V)}(X) } \\
            &\qquad = \big\{ \Set{1,2},\Set{1,5},\Set{1,2,5},\Set{1,3,4,5} \big\},
        \end{aligned} \nonumber
     \end{equation}
     being all complimentary subsets in the asymptotic model and
     \begin{equation}
        \begin{aligned}
            &\Set{X \subset V \colon |X| > 1, H(V) - H(X) + \RNCO(X) \leq \RNCO(V)} \\
            & \qquad = \Set{ X \subset V \colon |X|>1, \FuHash{\RNCO(V)}(X) = \FuHashHat{\RNCO(V)}(X) } \\
            & \qquad = \big\{ \Set{1,2},\Set{1,4},\Set{1,5},\Set{2,4},\Set{2,5},\Set{1,2,4},\\
            & \qquad\qquad \Set{1,2,5},\Set{1,3,4},\Set{1,3,5},\Set{1,4,5},\Set{2,3,4},\\
            & \qquad\qquad \Set{2,3,5},\Set{2,4,5},\Set{1,2,3,4},\Set{1,2,3,5},\\
            & \qquad\qquad \Set{1,2,4,5},\Set{1,3,4,5},\Set{2,3,4,5} \big\},
        \end{aligned} \nonumber
     \end{equation}
     being all complimentary subsets in the non-asymptotic model.
\end{example}

Then, the task reduces to finding a subset $X$ such that $\FuHash{\alpha}(X) = \FuHashHat{\alpha}(X)$, where $\alpha=\RACO(V)$ and $\alpha=\RNCO(V)$ for the asymptotic and non-asymptotic models, respectively. It can be converted to a minimization problem
               \begin{equation} \label{eq:SFMAlgo}
                    \min_{X \subseteq V_i \colon i \in X} \Set{ \FuHash{\alpha}(X) - r(X) } ,
                \end{equation}
where $i \in V$ and $V_i = \Set{1,\dotsc,i}$, based on which, we propose an algorithm for searching for the complimentary subset for SO (CompSetSO) in Algorithm~\ref{algo:CompSetSO}.

        \begin{algorithm} [t]
           \caption{complimentary subset for SO (CompSetSO)}
	       \label{algo:CompSetSO}
	       \small
	       \SetAlgoLined
	       \SetKwInOut{Input}{input}\SetKwInOut{Output}{output}
	       \SetKwFor{For}{for}{do}{endfor}
            \SetKwRepeat{Repeat}{repeat}{until}
            \SetKwIF{If}{ElseIf}{Else}{if}{then}{else if}{else}{endif}
	       \BlankLine
           \Input{the ground set $V$, an oracle that returns the value of $H(X)$ for a given $X \subseteq V$ and $\alpha$, which is determined based on Theorem~\ref{theo:algo:CompSetSO} or Theorem~\ref{theo:algo:CompSetSOExt}.}
	       \Output{$\hat{X}$ which is a complimentary user subset for SO}
	       \BlankLine
            $ r_{1} \leftarrow \FuHash{\alpha}(\Set{1})$ and $r_i \leftarrow \alpha - H(V), \forall i \in V \setminus \Set{1}$\;
            \For{$i=2$ \emph{\KwTo} $|V|$}{
                \eIf{there exists a non-singleton minimizer $\hat{X}$ of $\min_{X \subseteq V_i \colon i \in X} \Set{ \FuHash{\alpha}(X) - r(X) }$ such that $\hat{X} \subset V$}{
                    terminate iteration and return $\hat{X}$;}{
                    $ r_i \leftarrow r_i + \min_{X \subseteq V_i \colon i \in X} \Set{ \FuHash{\alpha}(X) - r(X)}$\; }
            }
	   \end{algorithm}

\begin{theorem} \label{theo:algo:CompSetSO}
    For the CompSetSO algorithm in Algorithm~\ref{algo:CompSetSO}, the output $\hat{X}$ is a complimentary user subset for the asymptotic and non-asymptotic models if the input $\alpha = \RACO(V)$ and $\alpha = \RNCO(V)$, respectively; If there is no output, there does not exist a complimentary subset for SO. \hfill\IEEEQED
\end{theorem}

The proof of Theorem~\ref{theo:algo:CompSetSO} is in Appendix~\ref{app:theo:algo:CompSetSO}.

\begin{example}  \label{ex:auxCompSetSOMinSumRate}
    We apply the CompSetSO algorithm to the asymptotic model of the system in Example~\ref{ex:aux} by inputting $\alpha=\RACO(V) = \frac{13}{2}$. It can be shown that for $i = 2$, $\hat{X} = \Set{1,2}$ is returned as a complimentary subset. For the non-asymptotic model, by inputting $\alpha = \RNCO(V) = 7$, we get $\hat{X} = \Set{1,2}$ returned as a complimentary subset for $i = 2$.
\end{example}

\subsubsection{Sufficient Condition}

In Theorem~\ref{theo:algo:CompSetSO}, knowing the value of the minimum sum-rate, $\RACO(V)$ or $\RNCO(V)$, is a prerequisite. However, if we obtain the value of $\RACO(V)$ or $\RNCO(V)$, say, by the modified decomposition algorithm in \cite{Ding2016IT} or the deterministic algorithms in \cite{CourtIT2014,MiloIT2015}, we necessarily know an optimal rate vector in $\RRACO(V)$ or $\RRNCO(V)$ for the global omniscience, in which case solving a local omniscience problem in a user subset may not be necessary. It is also not consistent with the advantage of SO that the local omniscience problem is less complex than the global one.
So, the question is: Can we find a complimentary user subset for SO without knowing the minimum sum-rate? The answer is yes. In this section, we derive the sufficient conditions for a subset to be complimentary such that the value of $\alpha$ in the CompSetSO algorithm can be relaxed from the exact value of $\RACO(V)$ or $\RNCO(V)$ to a lower bound on $\RACO(V)$ or $\RNCO(V)$ that can be obtained in $O(|V|)$ time.

\begin{lemma}\label{lemma:SOEQV2}
    In an asymptotic model, a non-singleton user subset $X \subset V$ is complimentary if $\FuHash{\alpha}(X) = \FuHashHat{\alpha}(X)$ for $\alpha = \sum_{i \in V} \frac{H(X) - H(\Set{i})}{|V| - 1}$; In a non-asymptotic model, a non-singleton user subset $X \subset V$ is complimentary if $\FuHash{\alpha}(X) = \FuHashHat{\alpha}(X)$ for $\alpha = \big\lceil \sum_{i \in V} \frac{H(X) - H(\Set{i})}{|V| - 1} \big\rceil$. \hfill \IEEEQED
\end{lemma}
The proof of Lemma~\ref{lemma:SOEQV2} is in Appendix~\ref{app:lemma:SOEQV2}. The values of $\alpha$ in Lemma~\ref{lemma:SOEQV2} are the lower bounds on $\RACO(V)$ and $\RNCO(V)$ for asymptotic and non-asymptotic models, respectively \cite[Theorem IV.1]{Ding2016IT}.

\begin{theorem} \label{theo:algo:CompSetSOExt}
    The CompSetSO algorithm in Algorithm~\ref{algo:CompSetSO} returns a complimentary user subset $\hat{X}$ for the asymptotic and non-asymptotic models if the input $\alpha = \sum_{i \in V} \frac{H(V) - H(\Set{i})}{|V| - 1}$ and $\alpha = \big\lceil \sum_{i \in V} \frac{H(V) - H(\Set{i})}{|V| - 1} \big\rceil$, respectively; If there is no output, there does not exist complimentary subset for SO.  \hfill\IEEEQED
\end{theorem}

The proof of Theorem~\ref{theo:algo:CompSetSOExt} is in Appendix~\ref{app:theo:algo:CompSetSOExt}. The case when there is no complimentary subset in both Theorems~\ref{theo:algo:CompSetSO} and \ref{theo:algo:CompSetSOExt} is given in Appendix~\ref{app:NoCompSetSO}.

\begin{example} \label{ex:auxCompSetSOLB}
    For the asymptotic model of the system in Example~\ref{ex:aux}, by inputting $\alpha = \sum_{i \in V} \frac{H(X) - H(\Set{i})}{|V| - 1} = \frac{23}{4}$ in the CompSetSO algorithm, we get output $\hat{X} = \Set{1,2}$ as a complimentary subset. By inputting $\alpha = \big\lceil \sum_{i \in V} \frac{H(X) - H(\Set{i})}{|V| - 1} \big\rceil = 6$ in the CompSetSO algorithm, we get output $\hat{X} = \Set{1,2}$ as a complimentary subset for the non-asymptotic model.
\end{example}

\subsubsection{Complexity}

In the CompSetSO algorithm, \eqref{eq:SFMAlgo} is a submodular function minimization (SFM) problem,\footnote{The submodularity of \eqref{eq:SFMAlgo} is proved in \cite{Ding2016IT} based on the submodularity of the entropy function $H$.} which can be solved in $O(\SFM(|V|))$ time. Let $\alpha$ be initiated according to Theorem~\ref{theo:algo:CompSetSOExt}. Then, in an asymptotic or non-asymptotic model, if there exists a complimentary user subset $\hat{X}$ for SO, it can be found in $O(|V| \cdot \SFM(|V|))$ time. Consider the situation when there is no complimentary user subset for SO. The CompSetSO algorithm completes in $O(|V| \cdot \SFM(|V|))$ time without any output. But, in this case, according to \cite[Theorem V.1]{Ding2016IT}, $\rv_V$ is updated to an optimal rate vector in $\RRACO(V)$ and $\RRNCO(V)$ for the asymptotic and non-asymptotic models, respectively. In summary, the CompSetSO algorithm determines either a complimentary user subset for SO or an optimal rate vector in $O(|V| \cdot \SFM(|V|))$ time.

\section{Multi-stage Successive Omniscience}

We show an example of multi-stage SO in a CCDE system.

\begin{example} \label{ex:auxMultiSO}
    Consider the system in Example~\ref{ex:aux} as a CCDE system where the linear combinations of packets are transmitted by random linear network coding (RLNC) scheme\cite{Ho2006RLNC}. For example, if $r_i = 2$, then user $i$ broadcasts the linear coding $S_i = \gamma^{\intercal} \Rz{i} = \sum_{j \in \Rz{i}} \gamma_j \RW{j} $ twice. Here, $\gamma = (\gamma_j : \RW{j} \in \Rz{i})$ and, at each broadcast, $\gamma_j$ is randomly chosen from a Galois field $\F_q$ with $q > H(V) \cdot |V|$. If $r_i = \frac{3}{2}$, then each packet $\RW{j} \in \RZ{i}$ is broken into two chunks: $\RW{j}^{(1)}$ and $\RW{j}^{(2)}$ and user $i$ broadcast $S_i = \sum_{j \in \RZ{i}} (\gamma_j^{(1)} \RW{j}^{(1)} + \gamma_j^{(2)} \RW{j}^{(2)})$ for six times. At each broadcast, each $\gamma_j^{(1)}$ and $\gamma_j^{(2)}$ are randomly chosen from a Galois field $\F_q$ with $q > 2 \cdot H(V) \cdot |V|$.

    For the asymptotic model, the global omniscience can be achieved by three stages of SO: By transmitting the rates $\rv'_V = (2,0,0,0,0)$, $\rv''_V = (2,0,0,0,1)$ and $\rv'''_V = (\frac{1}{2},0,\frac{1}{2},\frac{1}{2},0)$, the omniscience is achieved in $\Set{1,2}$, $\Set{1,2,5}$ and $\Set{1,2,3,4,5}$ in sequence. Here, we have $\rv'_V + \rv''_V + \rv'''_V = (\frac{9}{2},0,\frac{1}{2},\frac{1}{2},1) \in \RRACO(V)$, which also means that the optimal rate vector $(\frac{9}{2},0,\frac{1}{2},\frac{1}{2},1)$ can be implemented by three stages of SO. We have shown how to find the complimentary user subset $\Set{1,2}$ by the CompSetSO algorithm in Examples~\ref{ex:auxCompSetSOMinSumRate} and \ref{ex:auxCompSetSOLB}. The complimentary subset $\Set{1,2,5}$ is determined as follows.

    Since local omniscience in $\Set{1,2}$ is attained after the transmission of $\rv'_V$, we can treat $\Set{1,2}$ as a super user and assign a user index $12'$. For each user $i \in V \setminus \Set{1,2}$, we assign a new index $i'$ with $\RZ{i'} = \RZ{i} \cup \Gamma$, where $\Gamma$ contains all the transmissions that is received by user $i$ in the first stage of SO, i.e., all the broadcasts for achieving the local omniscience in $\Set{1,2}$. For the super user $12'$, we have $\RZ{12'} = \RZ{1} \cup \RZ{2}$. We construct the ground set of the new system as $V' = \Set{12',3',4',5'}$ with the observations $\RZ{i'}$ for all $i \in V'$. By applying the CompSetSO algorithm to the new system, we get $\Set{12',5'}$ as a complimentary subset, which corresponds to $\Set{1,2,5}$ in the original system.

    In the same way, for the non-asymptotic model, it can be shown that the global omniscience can be achieved by three stages of SO: By transmitting the rates $\rv'_V = (2,0,0,0,0)$, $\rv''_V = (3,0,0,1,0)$ and $\rv'''_V = (0,0,1,0,0)$, the omniscience is achieved in $\Set{1,2}$, $\Set{1,2,4}$ and $\Set{1,\dotsc,5}$ in sequence. And, $\rv'_V + \rv''_V + \rv'''_V = (5,0,1,1,0) \in \RRNCO(V)$, which also means that the optimal rate vector $(5,0,1,1,0)$ can be implemented by three stages of SO.
\end{example}

\section{Conclusion}
We studied the problem of how to efficiently search a complimentary user subset so that the omniscience of a discrete multiple random source among a set of users can be attained in a successive manner. Based on the existing necessary and sufficient condition for a user subset to be complimentary, we proposed a CompSetSO algorithm, which searches a complimentary subset for SO in both asymptotic and non-asymptotic models. We showed that inputting a lower bound, instead of the exact value, of the minimum sum-rate is sufficient for the CompSetSO algorithm to return either a complimentary subset or an optimal rate vector in $O(|V| \cdot \SFM(|V|))$ time. The CompSetSO algorithm can be implemented recursively so that the omniscience can be attained in multi-stages of SO.

For the future research work, it is worth studying how to implement the SO more efficiently than the CompSetSO algorithm. Also, for the CCDE problem, it would be of interest if the multi-stage SO can be implemented by network coding schemes other than RLNC.

\appendices

\section{Proof of Corollary~\ref{coro:SOEQV1}} \label{app:coro:SOEQV1}
    Based on Theorem~\ref{theo:SO}, we have $ \RACO(X) \leq \RACO(V) - H(V) + H(X) = \FuHash{\RACO(V)}(X) $ being the necessary and sufficient condition for $X$ to be complimentary in the asymptotic model.
    On the other hand, $\RACO(X) \geq \sum_{C \in \Pat} \frac{H(X) - H(C)}{|\Pat| - 1}, \forall \Pat \in \Pi'(X)$. Then,
    we have inequality $\sum_{C \in \Pat} \frac{H(X) - H(C)}{|\Pat| - 1} \leq \RACO(V) - H(V) + H(X), \forall \Pat \in \Pi'(X)$, which is equivalent to $\FuHash{\RACO(V)}(X) \leq \sum_{C \in \Pat} \FuHash{\RACO(V)}(C), \forall \Pat \in \Pi'(X)$, i.e., $\FuHash{\RACO(V)}(X) = \FuHashHat{\RACO(V)}(X)$. In the same way, we can prove that $X$ is complimentary in the non-asymptotic model if and only if $\FuHash{\RNCO(V)}(X) = \FuHashHat{\RNCO(V)}(X)$. \hfill\IEEEQED

\section{Proof of Theorem~\ref{theo:algo:CompSetSO}} \label{app:theo:algo:CompSetSO}

    For $\alpha = \RACO(V)$, let $\hat{X}$ be the user subset returned by the CompSetSO algorithm, i.e., we find a non-singleton user subset $\hat{X} \subset V$ that minimizes $\min\Set{ \FuHash{\alpha}(X) - r(X) \colon \hat{i} \in X \subseteq V_{\hat{i}}}$, for some $\hat{i} \in \Set{1,\dotsc,|V|}$. It also means that we have not found any non-singleton proper subset of $V$ that minimizes $\min\Set{ \FuHash{\alpha}(X) - r(X) \colon i \in X \subseteq V_i}$, i.e., $\Set{i}$ is the only minimizer, for all $i \in \Set{1,\dotsc,\hat{i}-1}$. In each iteration of the CompSetSO algorithm, we have $\rv_V \in P(\FuHash{\alpha},\leq)$, where $P(\FuHash{\alpha},\leq) = \Set{\rv_V \in \Real^{|V|} \colon r(X) \leq \FuHash{\alpha}(X)}$ is the polyhedron of $\FuHash{\alpha}$\cite{Ding2016IT}. Then, for all $X \subseteq V_{\hat{i}-1}$, we have $\sum_{i \in X} \FuHash{\alpha}(\Set{i}) = r(X) \leq \FuHash{\alpha}(X)$, i.e., $\FuHashHat{\alpha}(X) = \sum_{i \in X} \FuHash{\alpha}(\Set{i}) \leq \sum_{C \in \Pat} \FuHash{\alpha}(C)$ for all $\Pat \in \Pi(X)$.

    On the other hand, since $\hat{X} \subset V$ is a non-singleton minimizer of $\min\Set{ \FuHash{\alpha}(X) - r(X) \colon \hat{i} \in X \subseteq V_{\hat{i}}}$, we have $\FuHash{\alpha}(\hat{X}) - r(\hat{X}) \leq \FuHash{\alpha}(C) - r(C), \forall C \subseteq \hat{X} \colon \hat{i} \in C$. Then,
    \begin{equation}
        \begin{aligned}
            \FuHash{\alpha}(\hat{X}) & \leq \FuHash{\alpha}(C) + r(\hat{X} \setminus C)  \\
                                     & = \FuHash{\alpha}(C) + \sum_{i \in \hat{X} \setminus C} \FuHash{\alpha}(\Set{i}) \\ & \leq \FuHash{\alpha}(C) + \sum_{C' \in \Pat} \FuHash{\alpha}(C'),
        \end{aligned} \nonumber
    \end{equation}
    for all $C \subseteq \hat{X}$ such that $\hat{i} \in C$ and all $\Pat \in \Pi(\hat{X})$. Therefore, $\FuHash{\alpha}(\hat{X}) \leq \sum_{C \in \Pat} \FuHash{\alpha}(C)$ for all $\Pat \in \Pi(\hat{X})$. Then, $\FuHash{\alpha}(\hat{X}) = \FuHashHat{\alpha}(\hat{X})$. According to Corollary~\ref{coro:SOEQV1}. $\hat{X}$ is a complimentary user subset in the asymptotic model. In the same way, we can prove the statement for the non-asymptotic model when $\alpha=\RNCO(V)$. \hfill\IEEEQED

\section{Proof of Lemma~\ref{lemma:SOEQV2}} \label{app:lemma:SOEQV2}
    We have $0 \leq \alpha = \sum_{i \in V} \frac{H(X) - H(\Set{i})}{|V| - 1} \leq \RACO(V)$. If $\FuHash{\alpha}(X) = \FuHashHat{\alpha}(X)$, then
    $$ \sum_{C \in \Pat} \FuHash{\RACO(V)}(C) - \FuHash{\RACO(V)}(X) \geq \sum_{C \in \Pat} \FuHash{\alpha}(C) - \FuHash{\alpha}(X) \geq 0,$$
    for all $\Pat \in \Pi(X)$, i.e., $\FuHash{\RACO(V)}(X) = \FuHashHat{\RACO(V)}(X)$. According to Corollary~\ref{coro:SOEQV1}, $X$ is a complimentary subset in the asymptotic model. In the same way, we can prove that $X \subset V$ such that $|X| > 1$ is complimentary in the non-asymptotic model if $\FuHash{\alpha}(X) = \FuHashHat{\alpha}(X)$ for $\alpha = \big\lceil \sum_{i \in V} \frac{H(X) - H(\Set{i})}{|V| - 1} \big\rceil$.

\section{Proof of Theorem~\ref{theo:algo:CompSetSOExt}} \label{app:theo:algo:CompSetSOExt}
    According to Lemma~\ref{lemma:SOEQV2} and by using the same way as in the proof in Theorem~\ref{theo:algo:CompSetSO}, we can show that the output $\hat{X}$ is complimentary. For $\alpha = \sum_{i \in V} \frac{H(V) - H(\Set{i})}{|V| - 1}$, we have $\sum_{i \in V} H(\Set{i}) = |V| H(V) - (|V| - 1) \alpha$. If there is no output of the CompSetSO algorithm, it means
    \begin{equation}
        \begin{aligned}
            & \quad \sum_{i \in V} \FuHash{\alpha}(\Set{i}) - \sum_{C \in \Pat} \FuHash{\alpha}(C) \\
            & = \sum_{i \in V} \big( \alpha - H(V) + H(\Set{i}) \big) - \sum_{C \in \Pat} \big( \alpha - H(V) + H(C) \big) \\
            & = (|V| - |\Pat|)\big( \alpha - H(V) \big) + \sum_{i \in V} H(\Set{i}) - \sum_{C \in \Pat} H(C) \\
            & = \sum_{C \in \Pat} \big( H(V) - H(C) \big) -(|\Pat| - 1) \alpha < 0, \quad \forall \Pat \in \Pi'(V),
        \end{aligned} \nonumber
    \end{equation}
    which is equivalent to
    $$ \alpha > \sum_{C \in \Pat} \frac{H(V) - H(C)}{|\Pat| - 1}, \quad \forall \Pat \in \Pi'(V), $$
    i.e., $\alpha = \RACO(V)$. Also, we have
    \begin{equation}
        \begin{aligned}
            & \quad \FuHash{\alpha}(X) - \sum_{i \in X} \FuHash{\alpha}(\Set{i}) \\
            & = \alpha - H(V) + H(X) - \sum_{i \in X} \big( \alpha - H(V) + H(\Set{i}) \big) \\
            & = H(X) - \sum_{i \in X} H(\Set{i}) - (|X| - 1) \big( \alpha - H(V) \big)  \\
            & = \sum_{i \in X} \big( H(X) - H(\Set{i}) \big) - (|X| - 1) \big( \alpha - H(V) + H(X) \big)\\
            & > 0, \qquad\quad \forall X \subset V \colon |X| > 1,
        \end{aligned}\nonumber
    \end{equation}
    which is equivalent to
    \begin{equation}
        \begin{aligned}
            \RACO(V) & = \sum_{i \in V} \frac{H(X) - H(\Set{i})}{|X| - 1} \\
                     & > \RACO(V) - H(V) + H(X), \ \forall X \subset V \colon |X| > 1,
        \end{aligned}\nonumber
    \end{equation}
    i.e., there is no complimentary user subset for SO in the asymptotic model. In the same way, we can prove the statement for the non-asymptotic model. \hfill\IEEEQED

\section{Nonexistence of Complimentary Subset} \label{app:NoCompSetSO}
In a system, either asymptotic or non-asymptotic model, if there does not exist a complimentary subset for SO, it means the omniscience can not be attained with the minimum sum-rate in a manner such that the local omniscience in some non-singleton subset $X \subset V$ can be attained first. According to Corollary~\ref{coro:SOEQV1}, it happens when the $\Set{\Set{i} \colon i \in V}$ and $\Set{\Set{V}}$ are the only minimizers of $\FuHashHat{\alpha}(V) = \min_{\Pat \in \Pi(V)} \sum_{C \in \Pat} \FuHash{\alpha}(C)$, i.e., when $\Set{\Set{i} \colon i \in V}$ is the only maximizer of $\RACO(V) = \max_{\Pat \in \Pi'(V)} \sum_{C \in \Pat} \frac{H(V) - H(C)}{|\Pat|-1}$. This is the case when the components in $\RZ{V}$ are mutually independent. But, it is not necessary that a system with mutually independent $\RZ{V}$ does not have a complimentary subset.

\begin{example}
Consider the system where $V=\{1,2,3\}$ and each user observes respectively
    \begin{equation}
        \begin{aligned}
            \RZ{1} & = (\RW{a},\RW{b}),  \\
            \RZ{2} & = (\RW{b},\RW{c}),\\
            \RZ{3} & = (\RW{a},\RW{c}),
        \end{aligned}  \nonumber
    \end{equation}
where $\RW{j}$ is an independent uniformly distributed random bit. It can be shown that $\RACO(V) = \frac{3}{2}$, $\RRACO(V) = \Set{(\frac{1}{2},\frac{1}{2},\frac{1}{2})}$ and all the components in $\RZ{V}$ are mutually independent. In this case, $\Set{\Set{1},\Set{2},\Set{3}}$ and $\Set{\Set{1,2,3}}$ are the minimizers of $\min_{\Pat \in \Pi(V)} \sum_{C \in \Pat} \FuHash{\alpha}(C)$ and $\Set{\Set{1},\Set{2},\Set{3}}$ is the only maximizer of $\RACO(V) = \max_{\Pat \in \Pi'(V)} \sum_{C \in \Pat} \frac{H(V) - H(C)}{|\Pat|-1}$. In this system, there is no complimentary user subset for SO, i.e., the optimal rate vector $(\frac{1}{2},\frac{1}{2},\frac{1}{2})$ cannot be implemented in a way such that the local omniscience in some non-singleton proper subset of $V$ is attained before the global omniscience in $V$.

Consider the system where $V=\{1,2,3\}$ and each user observes respectively
    \begin{equation}
        \begin{aligned}
            \RZ{1} & = (\RW{a}),  \\
            \RZ{2} & = (\RW{b}),\\
            \RZ{3} & = (\RW{c}),
        \end{aligned}  \nonumber
    \end{equation}
where $\RW{j}$ is an independent uniformly distributed random bit. It can be shown that $\RACO(V) = 3$, $\RRACO(V) = \Set{(1,1,1)}$ and all the components in $\RZ{V}$ are mutually independent. In this case, all partitions $\Pat \in \Pi(V)$ are the minimizers of $\min_{\Pat \in \Pi(V)} \sum_{C \in \Pat} \FuHash{\alpha}(C)$ and all partitions in $\Pat \in \Pi'(V)$ are the maximizers of $\RACO(V) = \max_{\Pat \in \Pi'(V)} \sum_{C \in \Pat} \frac{H(V) - H(C)}{|\Pat|-1}$. In this system, all non-singleton subset $X \subset V$ are complimentary.
\end{example}

Based on the proof of Theorem~\ref{theo:algo:CompSetSOExt} in Appendix~\ref{app:theo:algo:CompSetSOExt}, when the CompSetSO algorithm does not return a complimentary user set for an asymptotic and non-asymptotic models when $\alpha = \sum_{i \in V} \frac{H(V) - H(\Set{i})}{|V| - 1}$ and $\alpha = \big\lceil \sum_{i \in V} \frac{H(V) - H(\Set{i})}{|V| - 1} \big\rceil$, respectively, we have $\alpha = \RACO(V)$ and $\alpha= \RNCO(V)$, which necessarily means that the omniscience cannot be attained in a successive manner.

\bibliography{ComplimentarySetBIB}

\begin{thebibliography}{10}
\providecommand{\url}[1]{#1}
\csname url@samestyle\endcsname
\providecommand{\newblock}{\relax}
\providecommand{\bibinfo}[2]{#2}
\providecommand{\BIBentrySTDinterwordspacing}{\spaceskip=0pt\relax}
\providecommand{\BIBentryALTinterwordstretchfactor}{4}
\providecommand{\BIBentryALTinterwordspacing}{\spaceskip=\fontdimen2\font plus
\BIBentryALTinterwordstretchfactor\fontdimen3\font minus
  \fontdimen4\font\relax}
\providecommand{\BIBforeignlanguage}[2]{{%
\expandafter\ifx\csname l@#1\endcsname\relax
\typeout{** WARNING: IEEEtran.bst: No hyphenation pattern has been}%
\typeout{** loaded for the language `#1'. Using the pattern for}%
\typeout{** the default language instead.}%
\else
\language=\csname l@#1\endcsname
\fi
#2}}
\providecommand{\BIBdecl}{\relax}
\BIBdecl

\bibitem{Csiszar2004}
I.~Csisz{\'a}r and P.~Narayan, ``Secrecy capacities for multiple terminals,''
  \emph{{IEEE} Trans. Inf. Theory}, vol.~50, no.~12, pp. 3047--3061, Dec. 2004.

\bibitem{Chan2011ITW}
C.~Chan, ``Linear perfect secret key agreement,'' in \emph{Proc. IEEE Inf.
  Theory Workshop}, Paraty, Brazil, 2011, pp. 723--726.

\bibitem{Roua2010}
S.~El~Rouayheb, A.~Sprintson, and P.~Sadeghi, ``On coding for cooperative data
  exchange,'' in \emph{Proc. IEEE Inf. Theory Workshop}, Cairo, Egypt, 2010,
  pp. 1--5.

\bibitem{SprintRand2010}
A.~Sprintson, P.~Sadeghi, G.~Booker, and S.~El~Rouayheb, ``A randomized
  algorithm and performance bounds for coded cooperative data exchange,'' in
  \emph{Proc. IEEE Int. Symp. Inf. Theory}, Austin, TX, 2010, pp. 1888--1892.

\bibitem{CourtIT2014}
T.~Courtade and R.~Wesel, ``Coded cooperative data exchange in multihop
  networks,'' \emph{{IEEE} Trans. Inf. Theory}, vol.~60, no.~2, pp. 1136--1158,
  Feb. 2014.

\bibitem{MiloIT2015}
N.~Milosavljevic, S.~Pawar, S.~E. Rouayheb, M.~Gastpar, and K.~Ramchandran,
  ``Efficient algorithms for the data exchange problem,'' \emph{{IEEE} Trans.
  Inf. Theory}, vol.~62, no.~4, pp. 1878 -- 1896, Feb. 2015.

\bibitem{ChanSuccessiveIT}
C.~Chan, Al-Bashabsheh, Q.~Zhou, N.~Ding, T.~Liu, and A.~Sprintson,
  ``Successive omniscience,'' \emph{{IEEE} Trans. Inf. Theory}, vol.~62, no.~6,
  pp. 3270--3289, Apr. 2016.

\bibitem{Ding2015NetCod}
N.~Ding, R.~A. Kennedy, and P.~Sadeghi, ``Iterative merging algorithm for
  cooperative data exchange,'' in \emph{Proc. Int. Symp. Network Coding},
  Sydney, Australia, 2015, pp. 41--45.

\bibitem{Ho2006RLNC}
T.~Ho, M.~Medard, R.~Koetter, D.~R. Karger, M.~Effros, J.~Shi, and B.~Leong,
  ``A random linear network coding approach to multicast,'' \emph{{IEEE} Trans.
  Inf. Theory}, vol.~52, no.~10, pp. 4413--4430, Oct. 2006.

\bibitem{ChanMMI}
C.~Chan, A.~Al-Bashabsheh, J.~Ebrahimi, T.~Kaced, and T.~Liu, ``Multivariate
  mutual information inspired by secret-key agreement,'' \emph{Proc. IEEE},
  vol. 103, no.~10, pp. 1883--1913, Oct. 2015.

\bibitem{Ding2016IT}
\BIBentryALTinterwordspacing
N.~Ding, C.~Chan, Q.~Zhou, R.~A. Kennedy, and P.~Sadeghi, ``Communication for
  omniscience,'' \emph{arXiv preprint arXiv:1611.08367}, 2016. [Online].
  Available: \url{https://arxiv.org/abs/1611.08367}
\BIBentrySTDinterwordspacing

\bibitem{Fujishige2005}
S.~Fujishige, \emph{Submodular functions and optimization}, 2nd~ed.\hskip 1em
  plus 0.5em minus 0.4em\relax Amsterdam, The Netherlands: Elsevier, 2005.

\bibitem{Court2011}
T.~Courtade and R.~Wesel, ``Weighted universal recovery, practical secrecy, and
  an efficient algorithm for solving both,'' in \emph{Proc. Annu. Allerton
  Conf. Commun., Control, and Comput.}, Monticello, IL, 2011, pp. 1349--1357.

\bibitem{Dilworth1944}
R.~P. Dilworth, ``Dependence relations in a semi-modular lattice,'' \emph{Duke
  Math. J.}, vol.~11, no.~3, pp. 575--587, Feb. 1944.

\end{thebibliography}

\end{document}